\documentclass
[preprint,prd,onecolumn,showpacs,showkeys,nobibnotes,titlepage,nofootinbib]{revtex4}%
\usepackage{amsfonts}
\usepackage{amsmath}
\usepackage{amssymb}
\usepackage{graphicx}
\usepackage{float}
\usepackage{rotating}%
\providecommand{\U}[1]{\protect\rule{.1in}{.1in}}

\ifx\pdfoutput\relax\let\pdfoutput=\undefined\fi
\newcount\msipdfoutput
\ifx\pdfoutput\undefined\else
\ifcase\pdfoutput\else
\ifx\paperwidth\undefined\else
\ifdim\paperheight=0pt\relax\else\pdfpageheight\paperheight\fi
\ifdim\paperwidth=0pt\relax\else\pdfpagewidth\paperwidth\fi
\fi\fi\fi
\begin{document}
\preprint{ }
\title[Fractional Mechanics]{Applications of Fractional Calculus to Newtonian Mechanics}
\author{Gabriele U. Varieschi}
\affiliation{Department of Physics, Loyola Marymount University - Los Angeles, CA 90045,
USA\footnote{Email: gvarieschi@lmu.edu}}
\affiliation{}
\author{}
\eid{ }
\eid{ }
\author{}
\affiliation{}

\begin{abstract}
We investigate some basic applications of Fractional Calculus (FC) to
Newtonian mechanics. After a brief review of FC, we consider a possible
generalization of Newton's second law of motion and apply it to the case of a
body subject to a constant force.

In our second application of FC to Newtonian gravity, we consider a
generalized fractional gravitational potential and derive the related circular
orbital velocities. This analysis might be used as a tool to model galactic
rotation curves, in view of the dark matter problem.

Both applications have a pedagogical value in connecting fractional calculus
to standard mechanics and can be used as a starting point for a more advanced
treatment of \textit{fractional mechanics}.

\end{abstract}
\startpage{1}
\endpage{ }
\maketitle

\section{\label{sect:introduction}Introduction}

Fractional Calculus (FC) is a natural generalization of calculus that studies
the possibility of computing derivatives and integrals of any real (or
complex) order (\cite{MR0361633}, \cite{MR1219954}, \cite{MR1658022}), i.e.,
not just of standard integer orders, such as first-derivative,
second-derivative, etc.

The history of FC started in 1695 when l'H\^{o}pital raised the question as to
the meaning of taking a fractional derivative such as $d^{1/2}y/dx^{1/2}$ and
Leibniz replied \cite{MR1219954}: \textquotedblleft...This is an apparent
paradox from which, one day, useful consequences will be
drawn.\textquotedblright\ 

Since then, eminent mathematicians such as Fourier, Abel, Liouville, Riemann,
Weyl, Riesz, and many others contributed to the field, but until lately FC has
played a negligible role in physics. However, in recent years, applications of
FC to physics have become more common (\cite{Herrmann:2011zza},
\cite{MR1890104}) in fields ranging from classical and quantum mechanics,
nuclear physics, hadron spectroscopy, and up to quantum field theory.

In theoretical physics we can now study the fractional equivalent of many
standard physics equations \cite{Herrmann:2011zza}: frictional forces,
harmonic oscillator, wave equations, Schr{\"{o}}dinger and Dirac equations,
and several others. In applied physics \cite{MR1890104}, FC methods can be
used in the description of chaotic systems and random walk problems, in
polymer material science, in biophysics, and other fields.

In this paper, we will review elementary definitions and methods of fractional
calculus and fractional differential equations. We will then apply these
concepts to some basic problems in Newtonian mechanics, such as possible
generalizations of Newton's second law of motion and applications of FC to
Newtonian gravity.

\section{\label{sect:fractional_calculus}Fractional calculus: a brief review}

Unlike standard calculus, there is no unique definition of derivation and
integration in FC. Historically, several different definitions were introduced
and used (for complete details see, for example, Refs. \cite{MR0361633} and
\cite{MR1219954}). All proposed definitions reduce to standard derivatives and
integrals for integer orders $n$, but they might not be fully equivalent for
non-integer orders of \textit{differ-integration}.\footnote{In FC derivation
and integration are often treated and defined as a single operation---with the
order $q$ respectively taken as a positive or negative real number---hence the
names \textit{differintegrals}, \textit{differintegration}, etc. Also, the
name \textit{fractional calculus} is actually a misnomer, since the order of
differintegration can be any real (or complex) number. A better name for this
field might be \textquotedblleft Differintegration to an arbitrary
order,\textquotedblright\ or similar.}

To gain an intuitive perspective of fractional derivatives
\cite{Herrmann:2011zza}, we consider some elementary functions such as the
exponential function $e^{kx\text{}}$, trigonometric functions $\sin(kx)$ or
$\cos(kx)$, and simple powers $x^{k}$, where $k$ is some constant. It is easy
to obtain recursive relations for derivatives of integer order $n$:%

\begin{gather}
\frac{d^{n}e^{kx}}{dx^{n}}=k^{n}e^{kx}\label{eq:2.1}\\
\frac{d^{n}\sin(kx)}{dx^{n}}=k^{n}\sin(kx+\frac{\pi}{2}n)\nonumber\\
\frac{d^{n}x^{k}}{dx^{n}}=\frac{k!}{(k-n)!}x^{k-n}.\nonumber
\end{gather}

These relations can be easily generalized to real or imaginary order $q$, with
appropriate gamma functions replacing the factorials when necessary:%

\begin{gather}
\frac{d^{q}e^{kx}}{dx^{q}}=k^{q}e^{kx};\ k\geq0\label{eq:2.2}\\
\frac{d^{q}\sin(kx)}{dx^{q}}=k^{q}\sin(kx+\frac{\pi}{2}q);\ k\geq0\nonumber\\
\frac{d^{q}x^{k}}{dx^{q}}=\frac{\Gamma\left(  k+1\right)  }{\Gamma
(k-q+1)}x^{k-q};\ x\geq0,\ k\neq-1,-2,...\nonumber
\end{gather}
with the functions restricted to $k\geq0$ and $x\geq0$ respectively, to ensure
the uniqueness of the above definitions.

These three approaches to fractional derivation were introduced respectively
by Liouville, Fourier, and Riemann and led to immediate generalizations for
analytic functions expanded in series of exponential, trigonometric, or power
functions. For example, the fractional derivative of a function $f(x)$,
according to Liouville, can be defined as%

\begin{gather}
f(x) =\sum\limits_{k =0}^{\infty}a_{k}e^{kx}\label{eq:2.3}\\
\frac{d^{q}f(x)}{dx^{q}} =\sum\limits_{k =0}^{\infty}a_{k}k^{q}e^{kx}
.\nonumber
\end{gather}

Applying instead the Riemann definition of fractional derivatives for power
functions to the case of a constant $C$, we obtain:%

\begin{equation}
\frac{d^{q}C}{dx^{q}}=\frac{d^{q}(Cx^{0})}{dx^{q}}=\frac{Cx^{-q}}{\Gamma
(1-q)}, \label{eq:2.4}%
\end{equation}
i.e., the derivative of a constant is not equal to zero in FC, unless this
condition is assumed as an additional postulate as in the so-called Caputo
derivative \cite{Herrmann:2011zza}.

More general definitions of fractional differintegrals exist in the
literature, such as the Gr{\"{u}}nwald formula \cite{MR0361633}:%

\begin{equation}
\frac{d^{q}f}{\left[  d(x-a)\right]  ^{q}}=\lim_{N\rightarrow\infty}\left\{
\frac{\left[  \frac{x-a}{N}\right]  ^{-q}}{\Gamma(-q)}\sum_{j=0}^{N-1}%
\frac{\Gamma(j-q)}{\Gamma(j+1)}f\left(  x-j\left[  \frac{x-a}{N}\right]
\right)  \right\}  , \label{eq:2.5}%
\end{equation}
which involves only evaluations of the function itself and can be used for
both positive and negative values of $q$. Another general definition is the
Riemann-Liouville fractional integral \cite{MR0361633}:%

\begin{equation}
\frac{d^{q}f}{\left[  d(x-a)\right]  ^{q}}=\frac{1}{\Gamma(-q)}\int_{a}%
^{x}\left(  x-y\right)  ^{-q-1}f(y)dy\left(  q<0\right)  , \label{eq:2.6}%
\end{equation}
which can only be applied directly to fractional integration $\left(
q<0\right)  $, but can be extended to fractional differentiation by combining
it with integer-order derivatives. It is beyond the scope of this paper to
analyze these and other formulas of FC more thoroughly. Interested readers
will find complete mathematical details in all the references cited in this section.

\section{\label{sect:generalizing_newtonian}Generalizing Newtonian mechanics}

One-dimensional Newtonian mechanics for a point-particle of constant mass $m$
is based upon Newton's second law of motion, a second-order ordinary
differential equation:%

\begin{equation}
\frac{d^{2}x(t)}{dt^{2}}=\frac{F}{m}. \label{eq:3.1}%
\end{equation}

We can easily think of at least two possible ways of generalizing Newton's
second law using fractional calculus:

\begin{itemize}
\item Change the order of the time derivative in the left-hand-side of Eq.
(\ref{eq:3.1}) to an arbitrary number $q$. This is motivated by current
studies of FC applied to physics \cite{Herrmann:2011zza}, where second-order
classical wave equations, Schr{\"{o}}dinger and Dirac equations, and several
others are generalized to fractional order $q$.

\item Generalize the expression of the force (or force field) $F$ on the
right-hand-side of Eq. (\ref{eq:3.1}) to include differintegrals of arbitrary
order $q$. This is also routinely done in applications of FC to physics
\cite{Herrmann:2011zza}, by selecting fractional generalizations of standard
electromagnetic potentials, in order to analyze phenomena in nuclear physics,
hadron spectroscopy, and other fields.
\end{itemize}

In the next two sub-sections, we will consider examples of these possible generalizations.

\subsection{\label{sect:constant_force}Constant force motion}

As our first example, we generalize Eq. (\ref{eq:3.1}) by using derivatives of
arbitrary (real) order $q$ and by considering a constant force per unit mass
$f=F/m=const$:\footnote{We note that, in order to ensure the dimensional
correctness of Eq. (\ref{eq:3.2}), we would need to redefine force so that its
dimensions become $M\ L\ T^{-q}$. Alternatively, if the customary dimensions
of force are used, a constant time scale factor $t_{SC}$ should be introduced
in Eq. (\ref{eq:3.2}): $\frac{d^{q}x(t)}{dt^{q}}=\frac{F}{m}t_{SC}%
^{2-q}=f\ t_{SC}^{2-q}$. We have adopted the former solution in the
following.}%

\begin{equation}
\frac{d^{q}x(t)}{dt^{q}}=\frac{F}{m}=f. \label{eq:3.2}%
\end{equation}

The general solution of this (extraordinary) differential equation is
\cite{MR0361633}:%

\begin{gather}
x(t)=\frac{d^{-q}f}{dt^{-q}}+c_{1}t^{q-1}+c_{2}t^{q-2}+...+c_{l}%
t^{q-l}\label{eq:3.3}\\
=\frac{f\ t^{q}}{\Gamma(1+q)}+c_{1}t^{q-1}+c_{2}t^{q-2}+...+c_{l}%
t^{q-l}\nonumber\\
with\left\{
\begin{array}
[c]{c}%
0<q\leq l<q+1,\text{{}}if\ q>0\\
l=0,\ if\ q\leq0
\end{array}
\right\}  ,\nonumber
\end{gather}
having used also Eq. (\ref{eq:2.4}) for the fractional derivative of the
constant force per unit mass $f$. The $l$ constants of integration, $c_{1}$,
$c_{2}$, ... , $c_{l}$, can be determined from the $l$ initial conditions:
$x(t_{0})$, $x^{\prime}(t_{0})$, ... , $x^{(l-1)}(t_{0})$.

For example, choosing for simplicity's sake $t_{0}=1$, the constants of
integration are determined by a set of linear equations in matrix form
$\mathbf{Mc}=\mathbf{d}$, where $\mathbf{c}=(c_{j})$\ is the vector of the
integration constants, while the matrix $\mathbf{M}=(m_{ij})$ and the vector
$\mathbf{d}=(d_{i})$ are obtained as follows:%

\begin{gather}
m_{ij}=\left[  \prod\limits_{n=1}^{i-1}(q-j-n+1)\right]  _{i,j=1,2,...,l}%
\label{eq:3.4}\\
d_{i}=\left[  x^{i-1}(1)-\frac{f}{\Gamma(1+q)}\prod\limits_{n=0}%
^{i-2}(q-n)\right]  _{i=1,2,...,l}\nonumber
\end{gather}

If the initial conditions and the force per unit mass are simply set to unity,
i.e., $x(1)=x^{\prime}(1)=x^{\prime\prime}(1)=...=1$ and $f=1$, our general
solution in Eq. (\ref{eq:3.3}), with the integration constants computed using
Eq. (\ref{eq:3.4}), can be easily plotted for different values of the order
$q$, as shown in Fig. 1.

\begin{figure}[ptb]
\includegraphics[width=\textwidth]{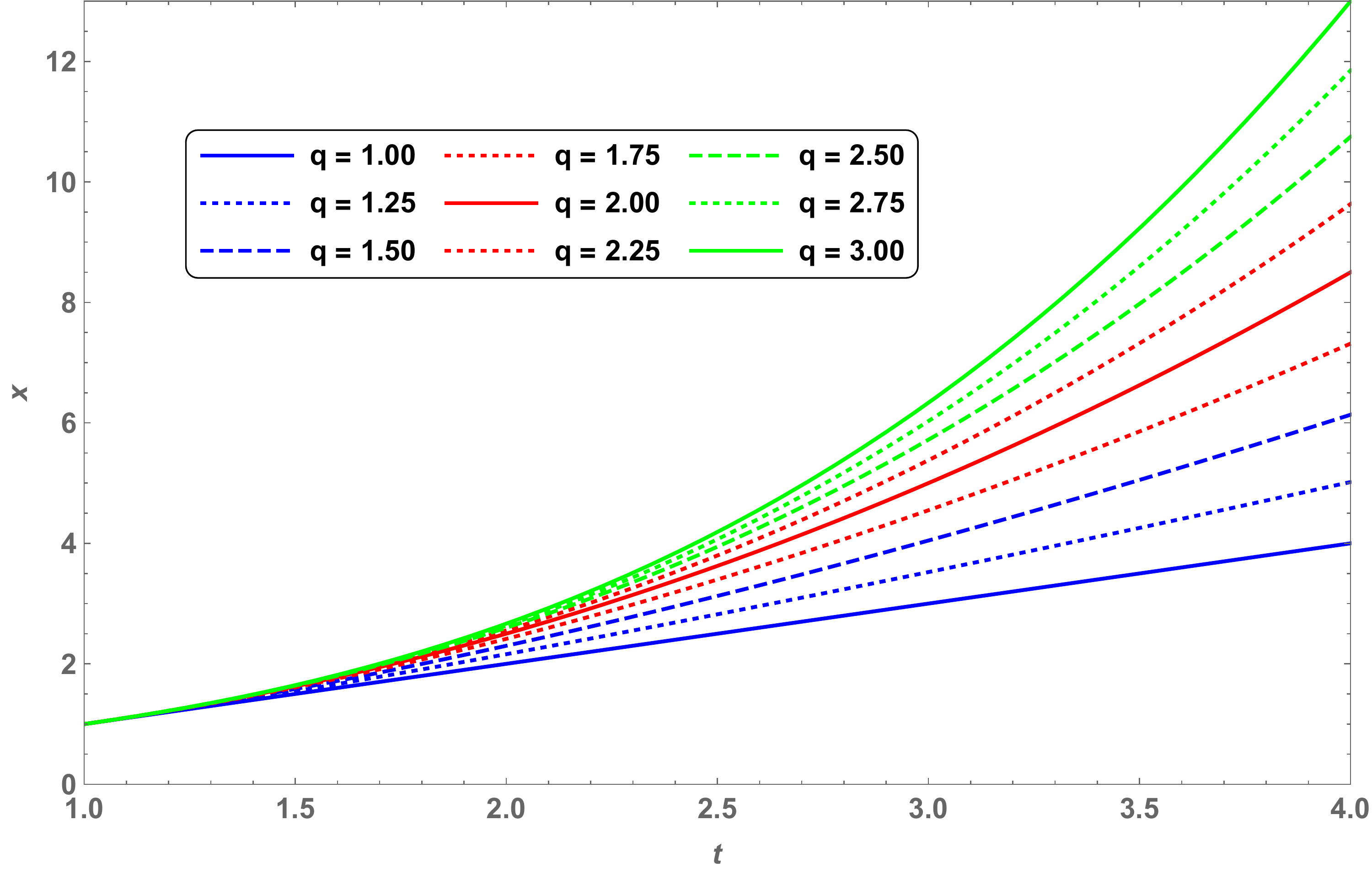}\caption{Position vs. time functions for a point-particle subject to a
constant force, using the generalized Newton's law in Eq. (\ref{eq:3.2}), with
the order $q$ ranging from $1$ to $3$, and with fractional increments. The
standard Newtonian solution is recovered for $q=2$ (red-solid curve), for a
motion with constant acceleration. The case for $q=1$ (blue-solid line)
represents a motion with constant velocity, while the $q=3$ case (green-solid
line) represents instead a motion with constant jerk. Other solutions (dashed
and dotted curves), for some fractional values of the order $q$, are also
shown in the figure.}%
\label{fig1}%
\end{figure}

This figure illustrates the resulting position vs. time functions for the
point-particle motion, subject to the generalized Newton's law in Eq.
(\ref{eq:3.2}), with the order $q$ ranging from $1$ to $3$ with fractional
increments. The standard Newtonian solution, $x(t)=\frac{1}{2}a\left(
t-t_{0}\right)  ^{2}+v_{0}\left(  t-t_{0}\right)  +x_{0}$, is obviously
recovered for $q=2$ (red-solid curve), for a motion with constant acceleration
$a=F/m$. Two other solutions for integer values of $q$ are presented: the case
for $q=1$ (blue-solid line) represents a simple motion with constant velocity
$v=F/m$: $x(t)=v\left(  t-t_{0}\right)  +x_{0}$; the $q=3$ case (green-solid
line) represents instead a motion with constant \textit{jerk}\footnote{The
higher-order derivatives of the position vs. time function (beyond the second
order) are usually called \textit{jerk} (3rd order), \textit{snap} or
\textit{jounce} (4th order), \textit{crackle} (5th order), \textit{pop} (6th
order), etc.} $j=F/m$: $x(t)=\frac{1}{6}j\left(  t-t_{0}\right)  ^{3}+\frac
{1}{2}a_{0}\left(  t-t_{0}\right)  ^{2}+v_{0}\left(  t-t_{0}\right)  +x_{0}$.

In Fig. 1, we also show (dashed and dotted curves) the position vs. time
functions for some fractional values of the order $q$ in Eq. (\ref{eq:3.2}).
These additional curves interpolate well between the integer-order functions
described above, showing that \textquotedblleft fractional
mechanics\textquotedblright\ would simply yield solutions for the motion of
the point-particle which are somewhat in-between the integer-order\ solutions.

One can't help but wonder what would the universe be like, if the fundamental
Newtonian (second) law of motion were based on an order $q$ different from the
standard value of two: for $q=1$, we would have a situation reminiscent of
Aristotelian physics, where a constant applied force would only achieve a
motion with constant velocity. For $q=3$, the application of a constant force
would yield a constant jerk (i.e., an acceleration changing at constant rate)
resulting in a motion much more difficult to control. Fractional values of $q$
would yield mechanical situations somewhat in-between those with integer $q$,
but the resulting dynamics would possibly be lacking of the other cardinal
principles of Newtonian mechanics, such as conservation laws or others.

\subsection{\label{sect:gravitational_force}Gravitational force}

Our second case of interest will be the generalization of Newton's law of
universal gravitation:%

\begin{equation}
\frac{d^{2}\overrightarrow{r}(t)}{dt^{2}}=\frac{\overrightarrow{F_{g}}}%
{m}=-\frac{GM}{r^{2}}\widehat{r}, \label{eq:3.5}%
\end{equation}
where $G$ is the universal gravitational constant, $M$ is the total mass of a
(spherically symmetric) source centered at the origin of a coordinate system,
$\widehat{r}$ and $r$ are respectively the radial unit vector and the radial
distance between the origin and\ a point-particle of mass $m$, subject to the
gravitational attraction.

In this case, we will modify the right-hand-side of Eq. (\ref{eq:3.5}) by
considering a generalized gravitational \textit{Riesz} potential
\cite{Herrmann:2011zza} $V_{RZ}$:%

\begin{equation}
V_{RZ}(\overrightarrow{r})=-\frac{G}{a}\int\nolimits_{%
\mathbb{R}
^{3}}\frac{dM}{\left(  s/a\right)  ^{q}}=-\frac{G}{a}\int\nolimits_{%
\mathbb{R}
^{3}}\frac{\rho(\overrightarrow{r^{\prime}})d^{3}\overrightarrow{r^{\prime}}%
}{\left(  \left\vert \overrightarrow{r}-\overrightarrow{r^{\prime}}\right\vert
/a\right)  ^{q}}, \label{eq:3.6}%
\end{equation}
where $s=\left\vert \overrightarrow{r}-\overrightarrow{r^{\prime}}\right\vert
$ is the distance between the infinitesimal source mass element $dM=\rho
(\overrightarrow{r^{\prime}})d^{3}\overrightarrow{r^{\prime}}$ and the
position$\overrightarrow{r}$ being considered. Due to the presence of the
fractional order $q$, a \textquotedblleft length scale\textquotedblright\ $a$
is needed to ensure the dimensional correctness of Eq. (\ref{eq:3.6}%
).\footnote{This approach \cite{Herrmann:2011zza} is based on a
3D-generalization of the convolution integral: $V(\overrightarrow{r}%
)=\int\nolimits_{%
\mathbb{R}
^{3}}\rho(\overrightarrow{r^{\prime}})w\left(  \left\vert \overrightarrow{r}%
-\overrightarrow{r^{\prime}}\right\vert \right)  d^{3}%
\overrightarrow{r^{\prime}}=\int\nolimits_{%
\mathbb{R}
^{3}}\frac{\rho(\overrightarrow{r^{\prime}})}{\left\vert \overrightarrow{r}%
-\overrightarrow{r^{\prime}}\right\vert }d^{3}\overrightarrow{r^{\prime}}$,
with a fractional weight function: $w\left(  \left\vert \overrightarrow{r}%
-\overrightarrow{r^{\prime}}\right\vert \right)  =$ $\frac{1}{\left\vert
\overrightarrow{r}-\overrightarrow{r^{\prime}}\right\vert ^{q}}$. The
resulting potential in Eq. (\ref{eq:3.6}) is equivalent to a 3D-version of the
\textit{Riesz fractional derivative}, which corresponds to a linear
combination of fractional Liouville integrals.}

For a spherical source of radius $R_{0}$ and uniform density $\rho
_{0}=M/\left(  \frac{4}{3}\pi R_{0}^{3}\right)  $,%

\begin{equation}
\rho(\overrightarrow{r^{\prime}})=\rho_{0}\ H(R_{0}-r^{\prime})=%
\begin{Bmatrix}
\rho_{0},\ for\ 0\leq r^{\prime}\leq R_{0}\\
0,\ for\ r^{\prime}>R_{0}%
\end{Bmatrix}
, \label{eq:3.7}%
\end{equation}
the integral in Eq. (\ref{eq:3.6}) can be evaluated analytically for any
(real) value of the fractional order $q$, inside and outside the source. In
general, we have \cite{Herrmann:2011zza}:%

\begin{align}
V_{RZ}(r)  &  =-\frac{G\rho_{0}}{r}\ \frac{2\pi a^{q-1}}{\left(  q-2\right)
\left(  q-3\right)  \left(  q-4\right)  }\label{eq:3.8}\\
&  \times\left\{
\begin{array}
[c]{c}%
\left[
\begin{array}
[c]{c}%
(r+R_{0})^{3-q}\left(  r-(3-q)R_{0}\right) \\
+\left(  R_{0}-r\right)  ^{3-q}\left(  r+(3-q)R_{0}\right)
\end{array}
\right]  ,\ for\ 0\leq r\leq R_{0}\\
\left[
\begin{array}
[c]{c}%
\left(  r+R_{0}\right)  ^{3-q}\left(  r-(3-q)R_{0}\right) \\
-\left(  r-R_{0}\right)  ^{3-q}\left(  r+(3-q)R_{0}\right)
\end{array}
\right]  ,\ for\ r>R_{0}%
\end{array}
\right\}  ,\nonumber
\end{align}
for the inner and outer solutions. The special cases for $q=2,3,4$ can be
obtained by recomputing the integrals for these particular values of $q$, or
by considering appropriate limits of $V_{RZ}(r)$, from the previous equation,
for $q\rightarrow2,3,4$. For example, for $q=2$, we obtain:%

\begin{equation}
V_{RZ}(r)|_{q=2}=-\frac{G\rho_{0}}{r}\pi a%
\begin{Bmatrix}
\left[  2rR_{0}+\left(  R_{0}^{2}-r^{2}\right)  \ln\left(  \frac{r+R_{0}%
}{R_{0}-r}\right)  \right]  ,\ for\ 0\leq r\leq R_{0}\\
\left[  2rR_{0}-\left(  r^{2}-R_{0}^{2}\right)  \ln\left(  \frac{r+R_{0}%
}{r-R_{0}}\right)  \right]  ,\ for\ r>R_{0}%
\end{Bmatrix}
. \label{eq:3.9}%
\end{equation}

Setting instead $q=1$ in Eq. (\ref{eq:3.8}), and using $\rho_{0}=M/\left(
\frac{4}{3}\pi R_{0}^{3}\right)  $, we recover the standard Newtonian potential:%

\begin{equation}
V_{RZ}(r)|_{q=1}=V_{Newtonian}(r)=%
\begin{Bmatrix}
-\frac{GM}{2R_{0}^{3}}\left(  3R_{0}^{2}-r^{2}\right)  ,\ for\ 0\leq r\leq
R_{0}\\
-\frac{GM}{r},\ for\ r>R_{0}%
\end{Bmatrix}
, \label{eq:3.10}%
\end{equation}
which yields the universal law of gravitation in Eq. (\ref{eq:3.5}), by using
just the outer potential from the last equation. Another very simple case is
the one for $q=0$, which yields a constant potential $V_{RZ}(r)|_{q=0}%
=-\frac{GM}{a}$, for both inner and outer solutions.

In Fig. 2, we illustrate the shape of these generalized gravitational Riesz
potentials following Eqs. (\ref{eq:3.8})-(\ref{eq:3.10}), for different values
of the fractional order $q$ ranging from zero to two. The $q=1$ case
(red-solid curve) represents the standard Newtonian gravitational potential.
All these plots were obtained by setting $G=M=R_{0}=a=1$ for simplicity's
sake, therefore the vertical grid line at $r=1.0$ in the figure denotes the
boundary between the inner ($0\leq r\leq R_{0}$) and the outer ($r>R_{0}$)
potentials. As already mentioned above, the $q=0$ case (blue-solid line)
corresponds to a constant potential $V_{RZ}(r)|_{q=0}=-\frac{GM}{a}$, while
the $q=2$ case (green-solid curve) is plotted using Eq. (\ref{eq:3.9}).

\begin{figure}[ptb]
\includegraphics[width=\textwidth]{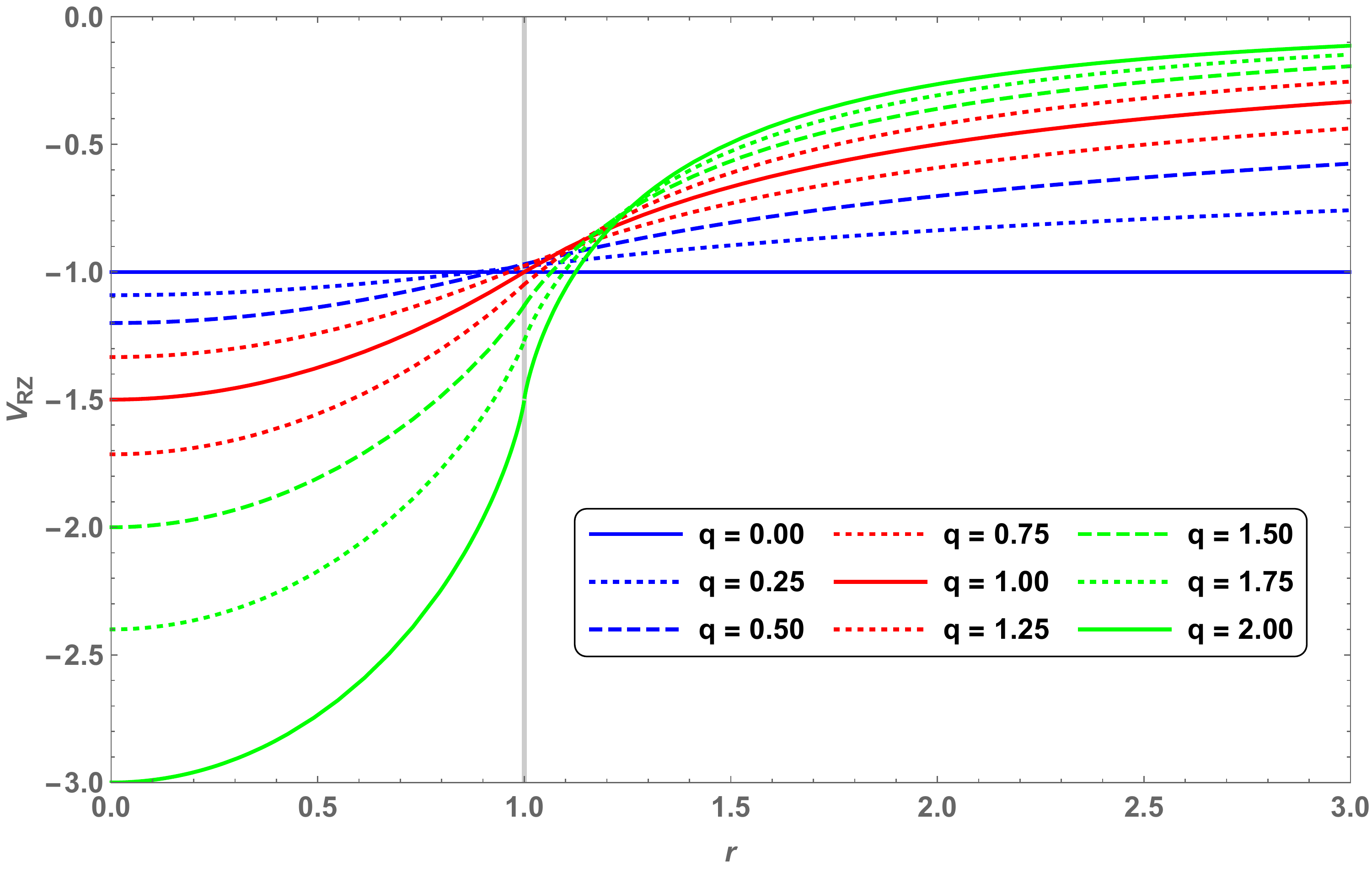}\caption{The generalized gravitational Riesz potentials, following Eqs.
(\ref{eq:3.8})-(\ref{eq:3.10}), for different values of the fractional order
$q$. The $q=1$ case (red-solid curve) represents the standard Newtonian
gravitational potential. We set $G=M=R_{0}=a=1$ for simplicity's sake, thus
the vertical grid line at $r=1.0$ in the figure denotes the boundary between
the inner ($0\leq r\leq R_{0}$) and the outer ($r>R_{0}$) potentials. The
$q=0$ case (blue-solid line) corresponds to a constant potential, while the
$q=2$ case (green-solid curve) is plotted using Eq. (\ref{eq:3.9}). Other
potentials (dashed and dotted curves), for some fractional values of the order
$q$, are also shown in the figure.}%
\label{fig2}%
\end{figure}

An interesting consequence of these generalized gravitational potentials is
the analysis of the resulting orbital circular velocities, for the inner and
outer solutions. From the generalized gravitational potentials in Eq.
(\ref{eq:3.8}), we can easily obtain the related gravitational force per unit mass:%

\begin{align}
\frac{\overrightarrow{F}_{RZ}(r)}{m}  &  =-\frac{dV_{RZ}(r)}{dr}%
\widehat{r}=-\frac{G\rho_{0}}{r^{2}}\widehat{r}\ \frac{2\pi a^{q-1}}{\left(
q-2\right)  \left(  q-3\right)  \left(  q-4\right)  }\label{eq:3.11}\\
&  \times\left\{
\begin{array}
[c]{c}%
\left[
\begin{array}
[c]{c}%
\left(  r+R_{0}\right)  ^{2-q}\left(  r-(3-q)R_{0}\right)  (R_{0}-(2-q)r)\\
+\left(  R_{0}-r\right)  ^{2-q}\left(  r+(3-q)R_{0}\right)  (R_{0}+(2-q)r)\\
-r(\left(  r+R_{0}\right)  ^{3-q}+\left(  R_{0}-r\right)  ^{3-q})
\end{array}
\right]  ,\ for\ 0\leq r\leq R_{0}\\
\left[
\begin{array}
[c]{c}%
\left(  r+R_{0}\right)  ^{2-q}\left(  r-(3-q)R_{0}\right)  (R_{0}-(2-q)r)\\
+\left(  r-R_{0}\right)  ^{2-q}\left(  r+(3-q)R_{0}\right)  (R_{0}+(2-q)r)\\
-r(\left(  r+R_{0}\right)  ^{3-q}-\left(  r-R_{0}\right)  ^{3-q})
\end{array}
\right]  ,\ for\ r>R_{0}%
\end{array}
\right\}  ,\nonumber
\end{align}
from which we can obtain the orbital circular velocities:%

\begin{equation}
v_{circ}(r)=\sqrt{r\frac{\left\vert \overrightarrow{F}_{RZ}(r)\right\vert }%
{m}}=\sqrt{r\left\vert \frac{dV_{RZ}(r)}{dr}\right\vert }. \label{eq:3.12}%
\end{equation}

Fig. 3 shows the plots of these orbital circular velocities for the same
values of the fractional order $q$ used in Fig. 2, and also by setting
$G=M=R_{0}=a=1$ as done previously. The $q=1$ case (red-solid curve)
represents the standard Newtonian situation, with the circular velocity
$v_{circ}=\sqrt{\frac{GM}{R_{0}^{3}}}r\sim r$ for $0\leq r\leq R_{0}$, and
$v_{circ}=\sqrt{\frac{GM}{r}}\sim1/\sqrt{r}$ for $r\geq R_{0}$ (the vertical
grid line at $r=1.0$ in the figure represents the boundary between the inner
and outer regions).

The $q=0$ case (blue-solid line) would not yield any circular velocity because
it corresponds to a zero-force case. The $q=2$ case (green-solid curve) is
computed using the special potential in Eq. (\ref{eq:3.9}), while in all the
other (fractional) cases the velocity plots interpolate well between the
integer cases outlined above.

\begin{figure}[ptb]
\includegraphics[width=\textwidth]{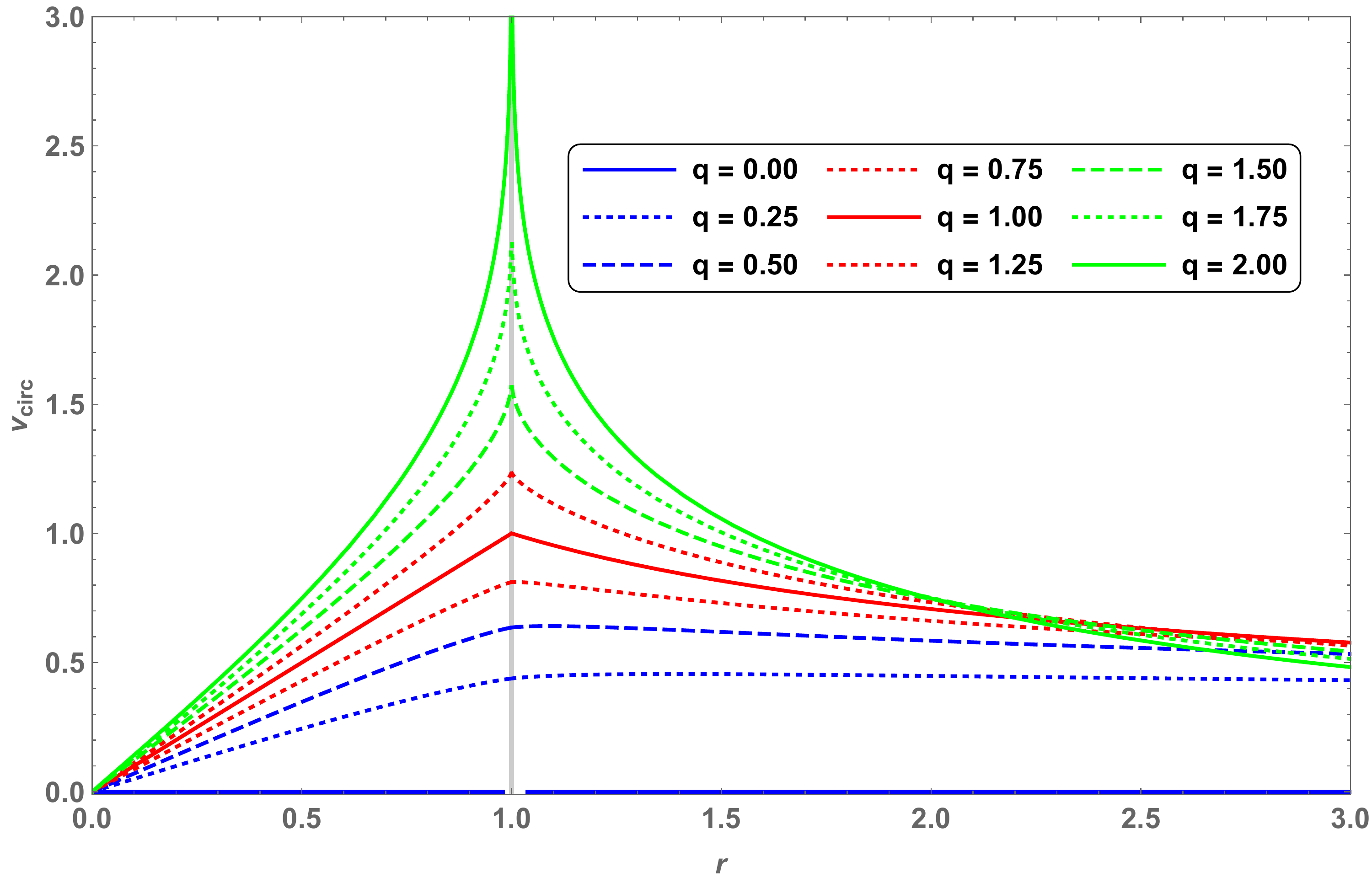}\caption{The orbital circular velocities, following Eqs. (\ref{eq:3.11}%
)-(\ref{eq:3.12}), for different values of the fractional order $q$. The $q=1$
case (red-solid curve) represents the standard Newtonian situation. Again, we
set $G=M=R_{0}=a=1$, thus the vertical grid line at $r=1.0$ in the figure
denotes the boundary between the inner ($0\leq r\leq R_{0}$) and the outer
($r>R_{0}$) velocities. The $q=0$ case (blue-solid line) corresponds to a zero
force situation, while the $q=2$ case (green-solid curve) is plotted using Eq.
(\ref{eq:3.9}) and (\ref{eq:3.12}). Other solutions (dashed and dotted
curves), for some fractional values of the order $q$, are also shown in the figure.}%
\label{fig3}%
\end{figure}

It is interesting to note that, for values of $q$ decreasing from one toward
zero, the rotational velocity curves in the outer ($r\geq R_{0}$) region show
a definite \textquotedblleft flattening\textquotedblright\ effect, which
becomes more pronounced for the lowest $q$ values (for example, in the
$q=0.25$ case, blue-dotted curve). This consideration might be of some
interest in relation with the well-established problem of \textit{dark matter}
in galaxies, as evidenced by the galactic rotation curves and their lack of
Newtonian behavior in the outer regions.

It is beyond the scope of this paper to perform any fitting of galactic
rotational curves, by means of our fractional model of the Riesz gravitational
potentials. However, it is interesting to note that the main feature of the
observed galactic rotational curves, i.e., their conspicuous flatness at
larger distances could be actually recovered for values of the fractional
order $q$ close to zero.

We also recall that one of the most popular alternative gravitational models,
Modified Newtonian Dynamics (MOND) (\cite{Milgrom:1983ca},
\cite{Milgrom:2001ny}), originated from a simple modification of Newton's
second law, to account for the observed properties of galactic motion. The
MOND modification can be applied to either side of Newton's second law
\cite{Milgrom:2001ny}: by setting the force to be proportional to a certain
function of the acceleration, or alternatively by changing the dependence of
the gravitational force on the distance. In this work we have shown that
similar modifications to the dynamics of a body in motion can also be obtained
by means of fractional calculus. Also, given possible connections between
fractional calculus and fractal geometry (\cite{MR1329030},
\cite{Nottale:2008ai}, \cite{Nottale:2011zz}, \cite{Calcagni:2016xtk},
\cite{Calcagni:2016azd}), a fractional approach to mechanics might be useful
to analyze complex structures such as galaxies or similar.

\section{\label{sect:conclusions}Conclusions}

In this work, we have applied fractional calculus to some elementary problems
in standard Newtonian mechanics. The main goal was to show that FC\ can be
used as a pedagogical tool, even in introductory physics courses, to gain more
insight into basic concepts of physics, such as Newton's laws of motion and
universal gravitation.

An intriguing consequence of FC, in connection with gravitational physics, is
the possibility of applying \textit{fractional mechanics }to the problem of
galactic rotation curves. We will leave to further studies to investigate in
more detail a possible connection between \textit{fractional mechanics} and
the dark matter puzzle.

\begin{acknowledgments}
This work was supported by a grant from the Frank R. Seaver College of Science
and Engineering, Loyola Marymount University, Los Angeles.
\end{acknowledgments}

\bibliographystyle{apsrev}
\bibliography{FRACTIONAL}

\end{document}